\documentclass[twocolumn,showpacs,preprintnumbers,amsmath,amssymb]{revtex4} 

\begin{document}
\title{Non-Poisson dichotomous noise: higher-order correlation 
functions and aging}
\author{P. Allegrini$^{1}$, Paolo Grigolini$^{2,3,4}$,Luigi
Palatella$^{3}$, Bruce J. West$^{5}$}
\affiliation{$^{1}$Istituto di 
Linguistica Computazionale del CNR,
Area della Ricerca di Pisa,\\ Via Moruzzi 1, 
56124, Ghezzano-Pisa, Italy }
\affiliation{$^{2}$Center for Nonlinear 
Science, University of North Texas,\\
P.O. Box 311427, Denton, Texas 76203-1427 }
\affiliation{$^{3}$Dipartimento di Fisica dell'Universit\`{a} di Pisa and INFM \\
Via Buonarroti 2, 56127 Pisa, Italy }
\affiliation{$^{4}$Istituto dei Processi Chimico Fisici del CNR,
Area della Ricerca di Pisa,\\ Via Moruzzi 1, 
56124, Ghezzano-Pisa, Italy}
\affiliation{$^{5}$Mathematics Division, Army Research 
Office, Research Triangle Park, NC 27709, USA}
\begin{abstract}
We study a two-state symmetric noise, with a given waiting time distribution 
$\psi (\tau )$, and focus our attention on the connection between the
four-time and the two-time correlation functions. The transition of $\psi
(\tau )$ from the exponential to the non-exponential condition yields the
breakdown of the usual factorization condition of high-order correlation
functions, as well as the birth of aging effects. We discuss the subtle
connections between these two properties, and establish the condition that
the Liouville-like approach has to satisfy in order to produce a correct
description of the resulting diffusion process.
\end{abstract}
\maketitle
%
%
%
%
%
%
%
%
%
%
%
%
\section{Introduction}

Dichotomous noise is one of the fundamental representations of stochastic
processes. It is used in random walks, quantum two-state systems, as well as
other mathematical models of physical and biological processes. This
representation is used because it is simple enough to obtain analytic
solutions to dynamical equations, yet rich enough to model a variety of
complex physical and biological phenomena. The history of such two-state
stochastic processes dates back more than a century to Markov representations
of random telegraphic signals and yet such noise still finds application in
models of contemporary complex phenomena. A few recent examples of complex
phenomena modeled by dichotomous stochastic processes are disorder-induced
spatial patterns \cite{buceta}; first-passage \cite{porra} and thermally
activated escape \cite{reineker} processes; hypersensitive transport \cite
{katja1}; rocking rachets \cite{katja2}; intermittent fluorescence \cite
{dahan}; stochastic resonance \cite{kapral,fulinski2,imkeller}; quantum
multifractality \cite{oliveira}; and blinking quantum dots \cite
{jung,neuhaser}. These and many other applications study the physical
effects of dichotomous fluctuations, either Poisson or non-Poisson, without
addressing, however, the consequences that relaxing the Poisson assumption
might have on the high-order correlation functions.

In this paper we are interested in the high-order correlation properties of
the dichotomous noise $\xi (t)$, that is, a symmetrical two-state
statistical process with the values +W and -W. Usually, for the purpose of
making statistical calculations we focus on stationary noise and use the
stationary correlation function, 
\begin{equation}
\Phi _{\xi }(|t_{1}-t_{2}|)=\frac{\langle \xi (t_{1})\xi (t_{2})\rangle }{\langle \xi ^{2}\rangle },
\end{equation}
where the brackets denote an average over an ensemble of realizations of the
dichotomous noise. It is worth illustrating the difference between this
dichotomous noise and a Gaussian noise with the same two-point correlation
function. The difference between the two processes resides in the high-order
correlation functions. Futhermore, because the noise is symmetric we only
need to focus on even-time correlation functions. According to Ref. \cite
{kubo}, for Gaussian noise the fourth-order correlation function is related
to the second-order correlation function via the following expression: 
\begin{equation}
\langle \xi (t_{1})\xi (t_{2})\xi (t_{3})\xi (t_{4})\rangle =\langle \xi (t_{1})\xi (t_{2})\rangle \langle \xi
(t_{3})\xi (t_{4})\rangle   \label{4pt}
\end{equation}
\[
+\langle \xi (t_{1})\xi (t_{3})\rangle \langle \xi (t_{2})\xi (t_{4})\rangle +\langle \xi (t_{1})\xi
(t_{4})\rangle \langle \xi (t_{2})\xi (t_{3})\rangle .
\]
The higher-order correlation functions are analogously defined. In the case
where all times are identical, the definition (\ref{4pt}) yields 
\begin{equation}
\langle \xi ^{2n}\rangle =(2n-1)!!\langle \xi ^{2}\rangle ^{n},  \label{gaussianequilibrium}
\end{equation}
a property ensuring that the distribution of $\xi $ is a Gaussian function.
By the same token, it seems natural to factor the fourth-order correlation
function for the dichotomous symmetric noise as 
\begin{equation}
\langle \xi (t_{1})\xi (t_{2})\xi (t_{3})\xi (t_{4})\rangle =\langle \xi (t_{1})\xi (t_{2})\rangle \langle \xi
(t_{3})\xi (t_{4})\rangle,   \label{factorization}
\end{equation}
with analogous prescriptions for the higher-order correlation functions. In
the case of equal times, the definition $\left( \ref{factorization}\right) $
reduces to
\begin{equation}
\langle \xi ^{2n}\rangle =\langle \xi ^{2}\rangle ^{n},  \label{4pt2}
\end{equation}
which is similar to, but not identical to $\left( \ref{gaussianequilibrium}%
\right) .$ Equation (\ref{4pt2}) is implied for the moments of a stochastic
process with the equilibrium distribution function 
\begin{equation}
p(\xi )=\frac{1}{2}[\delta (\xi -W)+\delta (\xi +W)].
\end{equation}
Hereafter, we refer to property (\ref{factorization}) and the
factorization of the corresponding
higher-order correlation equations, as Dichotomic Factorization (DF).

The vast majority of papers dealing with dichotomous noise assume the
statistics of the two-states to be Poisson, that is, the length of time the
system remains in a given state has a exponential distribution. It is
important to remark that the simplest physical phenomenon modeled by the
stochastic variable $\xi (t)$ is diffusion. This means that all the
properties of the phenomenon can be determined by the solution to the
stochastic equation 
\begin{equation}
\frac{dx}{dt}=\xi (t).  \label{langevin}
\end{equation}
Allegrini $et$ $al.$ \cite{allegrone} found that the evolution of the
probability density, corresponding to the dichotomous Langevin equation (\ref
{langevin}), is given by the Generalized Diffusion Equation (GDE) 
\begin{equation}
\frac{\partial p(x,t)}{\partial t}=\langle \xi ^{2}\rangle \int_{0}^{t}dt^{\prime }\Phi
_{\xi }(t-t^{\prime })\frac{\partial ^{2}}{\partial x^{2}}p(x,t^{\prime }),
\label{GDE}
\end{equation}
where the two-point correlation function under the integral is arbitrary.

It is interesting to note that the same GDE emerges from the analysis of C%
\'{a}ceres \cite{caceres}, who studied the Langevin equation 
\begin{equation}
\frac{dx}{dt}=-\gamma x(t)+\xi (t),  \label{friction}
\end{equation}
with $\xi (t)$ being a dichotomous noise and $\gamma $ a friction parameter
of arbitrary intensity. This same equation was studied in an earlier paper
by Annunziato \emph{et al.} \cite{annunziato}. It is evident that with $%
\gamma =0$ Eq. (\ref{friction}) becomes equivalent to Eq. (\ref{langevin}).
The equation for
densities found by C\'{a}ceres \cite{caceres} is identical to that found by
Annunziato \emph{et al.} and both results for $\gamma \rightarrow 0$ reduce
to Eq. (\ref{GDE}). These results are valid independently of the form of the
correlation function $\Phi _{\xi }(t)$. The fact that GDE is obtained using
these different approaches is significant, since the work by C\'{a}ceres
rests on van Kampen's lemma \cite{lemma} and the Bourret-Frisch-Pouquet
theorem \cite{bourret}, while the theory adopted by Annunziato \emph{et al.}
is the same as that used by Allegrini \textit{et al. }\cite{allegrone}, the
Zwanzig's projection method \cite{zwanzig}. In any event, both approaches
adopt of a Liouville-like perspective.

Bologna \emph{et al.} \cite{bolognone} established that the GDE produces the
same higher-order $x$-moments as those derived from the integration of the
diffusion equation, supplemented with the assumption that the correlation
functions of the dichotomous variable $\xi (t)$ fit the prescription
of DF. Bologna \emph{et al. }also established that the exact solution
of the GDE does not lead to the process of L\'{e}vy diffusion, a result
previously obtained using stochastic trajectories, thereby suggesting a possible
conflict between the adoption of stochastic trajectories obeying renewal
theory in the continuous time random walk (CTRW) formalism and the adoption
of a Liouville-like approach to the dynamics \cite{bolognone}. The DF
assumption is not explictly made by C\'{a}ceres
\cite{caceres}. However, the analysis of Bologna \emph{et
  al. }indicate that the theory of C\'{a}ceres \cite
{caceres} implies the DF property. Others have also assumed
non-Poisson statistics, while still retaining the DF property \cite{fulinski}.

We establish herein that the DF condition breaks down as a consequence of
the non-Poisson condition. Furthermore, we show that the violation of the DF
condition emerges from non-Poisson statistics in the same way as do aging
properties. These results have the desirable effect of establishing the
limits of validity of the elegant GDE, leaving aside for the present
the analysis of the issue as to whether the density and Liouville-like
formalism are compatible with the emergence of these properties.

\section{Four-time correlation function}

In this section we show that in the non-Poisson case, the four-time
correlation function of the dichotomous noise departs from the DF
prescription. It has to be pointed out that our arguments are based on
examining a single sequence $\xi $, and thus on time averages, rather than
on ensemble averages. We assume that the theoretical sequence is built
up 
by creating a sequence $\{\tau _{i}\}$ of real positive numbers using
the probability density 
\begin{equation}
\psi {(\tau )}=(\mu -1)\frac{T^{(\mu -1)}}{(\tau +T)^{\mu }}.
\label{tsallis}
\end{equation}
The choice of this analytical form is determined by simplicity, in which we
obtain in the time asymptotic limit an inverse power law with index $\mu $,
while satisfying the normalization condition 
\begin{equation}
\int_{0}^{\infty }\psi {(\tau )}d\tau =1.  \label{normalization}
\end{equation}
The parameter $T>0$ insures the normalization condition, required
by the fact that $\psi (\tau )$ is a probability density and is related to
the average time interval generated by the density. To generate a
realization of the time series we split the time axis into many time
intervals of lengths detemined by the set of numbers $\left\{ \tau
_{i}\right\} $. The first interval begins at time $t=0$ and ends at $t=\tau
_{1}$, the second begins at $t=\tau _{1}$ and ends at $t=\tau _{1}+\tau _{2}$%
, the third begins at $t=\tau _{1}+\tau _{2}$ and ends at $t=\tau _{1}+\tau
_{2}+\tau _{3}$, and so on. We refer to this sequence of time intervals,
which is not observable, as the theoretical sequence. The dichotomic
sequence under study in this paper, which can be observed, is created as
follows. At the beginning of any time interval we toss a coin, and fill
the interval with either the value $W$ or the value $-W$, according to
whether we get a head or a tail. Thus, if we move along the observable
sequence, we meet large time portions of the sequence, within which the
sequence retains the same value, either $W$ or $-W$. We refer to these time
intervals with the same value of $\xi $, as \emph{experimental} laminar
regions and to the corresponding distributions of time lengths as $\psi
_{exp}(\tau )$. The adoption of the suggestive term \emph{experimental}
reflects the fact that this procedure is the same as the one we would adopt
when making a real experimental observation. A relevant example is the
phenomenon of blinking quantum dots \cite{neuhaser}, which has been the
object of some very interesting theoretical papers \cite
{barkaisilbey,bouchaud} using dichotomous stochastic processes. A single
quantum dot undergoing a process of resonant fluorescence produces an
intermittent light signal, which can be identified with the sequence $\xi (t)
$ here under study, with $W$ and $-W$, meaning light-on and light-off,
respectively.

We point out that $\psi _{exp}(\tau )$ does not necessarily coincide
with $\psi (\tau )$. According to Ref. \cite{greatpaper} the theoretical waiting
time distribution $\psi (t)$ is connected to the experimental waiting time
distribution by the Laplace transform relation 
\begin{equation}
\hat{\psi}(u)=\frac{2\hat{\psi}_{exp}(u)}{1+\hat{\psi}_{exp}(u)},
\end{equation}
where the Laplace transform of a function $f\left( t\right) $ is denoted by $%
\hat{f}\left( u\right) $. However, in the time asymptotic
limit $\psi _{exp}(\tau )$ has the same inverse power law form as does $\psi
(\tau )$, that being Eq. (\ref{tsallis}), with the same power-law index $\mu 
$. In the special case of blinking quantum dots the experimental waiting
time distribution is found to be an inverse power law with index $\mu <2$. 
Here we consider the complementary case $\mu >2$, so as to realize a
condition compatible with the existence of a stationary correlation
function for $\xi (t)$.

Due to the theoretical prescription that we adopt to realize the dichotomic
sequence under study, a given experimental laminar region, namely, a time
interval where, as earlier pointed out, $\xi (t)$ keeps the same sign, might
correspond to an arbitrarily large number of theoretical time intervals, to
which the coin tossing procedure assigns the same sign. We shall refer to
these theoretical time intervals as theoretical laminar regions, or, more
simply, as laminar regions. It is evident that the beginning of a laminar
region corresponds to the occurrence of a random event, namely the coin
tossing that determines its sign. The laminar regions are not observable,
while the experimental laminar regions are observable, by definition, and
begin and end with a random event. We cannot establish if other random
events occur or not, and how many, between the beginning and the end of an
experimental laminar region.

The theoretical approach that we adopt in this section rests on the same
time average procedure as that adopted by Geisel \textit{et al.} \cite
{geisel}. Let us devote some attention to the prescription given by these
authors to evaluate the two-point correlation function $\Phi _{\xi
}(|t_{2}-t_{1}|)$ \cite{geisel}: 
\begin{equation}
\Phi _{\xi }(t_{2}-t_{1})=\frac{\int\limits_{t_{2}-t_{1}}^{\infty }[\tau
-(t_{2}-t_{1})]\psi (\tau )d\tau }{\int_{0}^{\infty }\tau \psi (\tau )d\tau }%
,  \label{geisel}
\end{equation}
where we assume $t_{2}>t_{1}.$ This equation for the correlation
function implies that, with a window of size $\Delta =t_{2}-t_{1}$ we move
along the entire (infinite) theoretical sequence of laminar regions and
count how many window positions are compatible with the window being located
within a theoretical laminar region, which must have a length larger than
the window size. In addition we have to count the total number of window
positions. In other words, the stationary correlation function of $\xi
\left( t\right) $ is nothing but the probability that the two times $t_{1}$
and $t_{2}$ are located within the same laminar region. If these two times
are located in different laminar regions, the adoption of the coin tossing
procedure for any contribution of a given sign to the correlation
function would produce, with equal probability, a contribution with opposite
sign, thereby providing a vanishing contribution. An attractive way to
explain this procedure is through the concept of random events. First of
all, the lengths of the laminar regions are determined by the random drawing
of the numbers $\tau $, with distribution $\psi (\tau )$. At the border
between one laminar region and the next we toss a coin to decide the sign of
the next laminar region. This coin tossing is a random event and no random
event can occur between two times located in the same laminar region. If the
two times are located in different laminar regions, one or more random
events must have occurred between them. Thus the correlation function $%
\Phi _{\xi }(|t_{1}-t_{2}|)$ can also be interpreted as the probability that
no random event occurs between times $t_{1}$ and $t_{2}$.

We evaluate the four-time correlation function, using the same arguments.
Consider four times, ordered as $t_{1}<t_{2}<t_{3}<t_{4}$. The corresponding
correlation function exists, under the following conditions. The first
condition is that all four times are located in the same laminar region. The
second condition is compatible with the pairs $(t_{1},t_{2})$ and $%
(t_{3},t_{4})$ being located in distinct laminar regions. This means that
the times $t_{1}$ and $t_{2}$ belong to a laminar region, denoted by $T_{1,2}
$, the times $t_{3}$ and $t_{4}$ belong to a laminar region denoted by 
$T_{3,4}$, and $T_{1,2}\neq T_{3,4}$. Using the random event concept, the
second condition implies that no random event occurs between $t_{1}$ and $%
t_{2}$, or between $t_{3}$ and $t_{4}$, while at least one random event
occurs between $t_{2}$ and $t_{3}$.

We use the notation $p(ij)$ to denote the probability that $t_{i}$ and $%
t_{j} $ belong to the same laminar region. Thus the prescription for the
correlation function given by Eq. (\ref{geisel}) can be expressed as the
probability function 
\begin{equation}
\Phi _{\xi }(t_{2}-t_{1})=p(12).
\label{correlationfunctioninbayesannotation}
\end{equation}
We also use the notation 
\begin{equation}
p\left( \overline{ij}\right) \equiv 1-p\left( ij\right)  \label{notation}
\end{equation}
to denote the probability that at least one transition occurs between times $%
t_{i}$ and $t_{j}$. It is convenient to use the conditional probability
concept, and the Bayesian notation (see, for instance, \cite{bayes}). We
denote the joint probability of events $A$ and $B$ by $p(A,B)$ and the
conditional probability of occurrence of event $A$ given event $B$ with $%
p(A|B)$. Thus, we have 
\begin{equation}
p(A|B)=\frac{p(A,B)}{p(B)}.  \label{cond}
\end{equation}
We denote the conditional probability that event $A$ occurs, given that
event $B$ does not, by $p(A|\overline{B})$. 
Using the prescription of Eq. (\ref{cond}), the latter conditional
probability, $p(A|\overline{B})$, is expressed as follows
\begin{equation}
p(A|\overline{B})=\frac{p(A)-p(A,B)}{1-p(B)},  \label{noevent}
\end{equation}
where we have used the relation $p\left( A\right) =p(A,B)+p(A,\overline{B})$ for
the numerator$.$

The probability that times $t_{i}$ and $t_{j}$ belong to the same laminar
region $T_{i,j}$ and that, simultaneously, times $t_{r}$ and $t_{s}$ belong
to the same laminar region $T_{r,s}$, regardless of whether $T_{i,j}$
coincides with $T_{r,s}$, or not, is a joint probability expressed by the
symbol $p(ij,rs)$. Thus the four-time correlation function can be formally
expressed as follows: 
\begin{equation}
\frac{\langle \xi (t_{1})\xi (t_{2})\xi (t_{3})\xi (t_{4})\rangle }{\langle
\xi ^{2}\rangle ^{2}}=p(12,34).  \label{4timestrajectories}
\end{equation}

On the other hand, using the notation introduced earlier, we have two
contributions to the four-time correlation function. The first
contributions is determined by all four times being in the same laminar
region with no random event occuring between $t_{1}$ and $t_{4}$ (condition
1), whereas the second contribution corresponds to the probability that at
least one random event occurs between $t_{2}$ and $t_{3}$, given the
condition that no random event occurs between $t_{1}$ and $t_{2}$ and none
between $t_{3}$ and $t_{4}$ (condition 2): 
\begin{equation}
p(12,34)=p(14)+p(\overline{23})p(12|\overline{23})p(34|\overline{23}).  \label{intermediate}
\end{equation}
Eq. (\ref{intermediate}) corresponds to the superposition of independent
contributions from condition 1 and condition 2. The contribution due to
condition 1, $p(14)$, according to the earlier definitions, is the
probability that $t_{1}$ and $t_{4}$ belong to the same laminar region. The
contribution due to condition 2 is given by the second term on the right
hand side of Eq. (\ref{intermediate}). Again, according to the notation that
we are using, see Eq.(\ref{notation}), $p(\overline{23})$ is the probability that
a random event occurs between $t_{2}$ and $t_{3},$ thereby disconnecting the
two laminar regions. Consequently $p(12|\overline{23})$ is the probability that $%
t_{1}$ and $t_{2}$ belong to the same laminar region given that at least one
random event occurs between $t_{2}$ and $t_{3}$. Finally, $p(34|\overline{23})$
is the probability that $t_{3}$ and $t_{4}$ belong to the same laminar
region, given that at least one random event occurs between $t_{2}$ and $%
t_{3}$. Thus, the product of these three probabilities is the appropriate
quantity corresponding to condition 2.

To transform the equality Eq. (\ref{intermediate}) into a relation involving
correlation functions, we use Eq.(\ref{4timestrajectories}), for the
four-time correlation function. The two-time correlation functions emerge
from the second term on the right hand side of Eq. (\ref{intermediate}) via
the proper use of Eq. (\ref{correlationfunctioninbayesannotation}), Eq. (\ref
{noevent}) and Eq. (\ref{notation}). Thus, we obtain 
\begin{equation}
\frac{\langle \xi (t_{1})\xi (t_{2})\xi (t_{3})\xi (t_{4})\rangle }{\langle
\xi ^{2}\rangle ^{2}}=\Phi _{\xi }(t_{4}-t_{1})  \label{centralresult}
\end{equation}
\[
+\frac{(\Phi _{\xi }(t_{2}-t_{1})-\Phi _{\xi }(t_{3}-t_{1}))(\Phi _{\xi
}(t_{4}-t_{3})-\Phi _{\xi }(t_{4}-t_{2}))}{1-\Phi _{\xi }(t_{3}-t_{2})}. 
\]
Eq. (\ref{centralresult}) is a major result, being an exact expression
for the four-time correlation function independently of the statistics of
the dichotomous process. We stress that the general form of Eq. (\ref
{centralresult}) is not factorable and is therefore distinct from DF.

Note that in the Poisson case, the waiting time distribution $\psi (t)$ is
exponential. Using the prescription given by Eq. (\ref{geisel}) it is not
difficult to show that the correlation function of $\xi $ is also
exponential. Then, after tedious but straightforward algebra, we establish
that Eq. (\ref{centralresult}) reduces to 
\begin{equation}
\langle \xi (t_{1})\xi (t_{2})\xi (t_{3})\xi (t_{4})\rangle =\langle \xi
(t_{1})\xi (t_{2})\rangle \langle \xi (t_{3})\xi (t_{4})\rangle ,
\end{equation}
which coincides with Eq.(\ref{factorization}), that is, the process becomes
compatible with the DF. Given that the DF holds true for the four-time
correlation function, it is possible to extend the DF property to the $2N$%
-time correlation function using induction.

Thus, we conclude that the four-time correlation condition (\ref
{centralresult}), for waiting times that have non-Poisson statistics,
violates the DF underlying Eq. (\ref{GDE}). This violation of the
factorization property seems to be a satisfactory explanation of why the GDE 
\cite{bolognone} does not yield the proper L\'{e}vy diffusion in the
asympotic limit. On the other hand, using the results of this section we
recover the results of the numerical calculations and theoretical prediction
of the fourth moments obtained by Allegrini \textit{et al. }\cite{jacopo}. To
establish this latter point we integrate Eq. (\ref{langevin}) with the initial
condition $x(0)$ for all the trajectories. Furthermore, we evaluate the
fourth power of $x(t)$, and average over all the trajectories of the Gibbs
ensemble. By using the stationary condition, which makes this correlation
function depend only on the time differences, rather than on the absolute
time, we obtain 
\begin{eqnarray}
&&\langle x^{4}(t)\rangle \label{4thmoment} \\
&&=8\int\limits_{0}^{t}dt_{4}\int\limits_{0}^{t_{4}}dt_{3}\int%
\limits_{0}^{t_{3}}dt_{2}\int\limits_{0}^{t_{2}}dt_{1}\langle \xi (t_{1})\xi
(t_{2})\xi (t_{3})\xi (t_{4})\rangle.  \nonumber 
\end{eqnarray}
Introducing the newly obtained expression for the fourth order correlation
Eq. (\ref{centralresult}) into (\ref{4thmoment}), in the time asymptotic
limit the leading contribution to the fourth moment is given by the first
term on the right hand side of Eq. (\ref{centralresult}). Therefore we
replace the integrand in Eq. (\ref{4thmoment}) with $\Phi _{\xi
}(t_{4}-t_{1})$, and using the inverse power-law form of the correlation
function, we carry out the four time integrations and obtain $\langle
x^{4}(t)\rangle \propto t^{6-\mu }$. By extending this way of proceeding
to the calculation of the $2n$-times correlation function, we derive the
general result 
\begin{equation}
\langle x^{2n}(t)\rangle \propto t^{2n-\mu+2 },  \label{higher}
\end{equation}
for $2\leq \mu \leq 3$, and $\langle x^{2n}(t)\rangle \propto t^{2n-1
 }$ for
$\mu >3$, in agreement with the numerical results of Ref. \cite{jacopo}.

The asymptotic result (\ref{higher}) establishes that the $2n$-moments do
not have the scaling corresponding to the DF condition. If we assume that the
condition of Eq. (\ref{factorization}) applies, in keeping with the nature
of the GDE, instead of (\ref{higher}) we would obtain $\langle
x^{2n}(t)\rangle \propto t^{2n(4-\mu) /2}$, with one factor of $\mu $
occuring for each order of the moment. Consequently, the DF implies the existence of the
scaling $x\propto t^{\delta }$, with the scaling index given by 
\begin{equation}
\delta =\frac{4-\mu }{2} \mbox{ for $2\leq \mu \leq 3$, }\delta=\frac{1}{2}\mbox{ for $\mu >3$, } \label{momentscaling}
\end{equation}
where $\mu -1$ is the L\'{e}vy index. This later result agrees with the
scaling predicted by the GDE, as established in Ref. \cite{bolognone}. Here
the central fact to keep in mind is that Eq. (\ref{langevin}) generates L%
\'{e}vy walks, rather than L\'{e}vy flights. A L\'{e}vy flight is a kind of
random walk in which the step lengths have an inverse power-law
distribution, so the second moment of the dynamical variable diverges. The L%
\'{e}vy walk, on the other hand, ties the length of a step to the time
required to take the step, resulting in a finite second moment for the
dynamical variable. Furthermore, it takes an infinite time for a L\'{e}vy
walk to yield the same scaling as a corresponding L\'{e}vy flight, the
latter scaling index being given by 
\begin{equation}
\delta =\frac{1}{\mu -1} \mbox{ for $2\leq \mu \leq 3$, }\delta=\frac{1}{2}\mbox{ for $\mu >3$} .
\end{equation}
For this reason, the L\'{e}vy walk, introduced by Shlesinger
\textit{et al. }%
\cite{west}, can be considered to be a manifestation of the Living State of
Matter (LSM) \cite{gerardone}, in the sense described in some recent work%
\cite{martina,buiatti}. The LSM is interpreted as the existence of a
scaling condition intermediate between that of dynamics and thermodynamics and
which can last forever. 

\section{Aging}

In this section we adopt the Bayesan formalism to evaluate the
correlation functions in a non-stationary condition. This enables us to
establish that the breakdown of the DF condition is closely related to aging.

Before proceeding with the formalism, we briefly review why non-Poisson
statistics produces aging, as discussed in detail in Refs. \cite
{gerardone,barkai}. Suppose that we create an infinite sequence of time
intervals of length $\tau _{i}$, namely, the theoretical sequence discussed
earlier. As mentioned, we create the observable sequence by filling the time
intervals, called laminar regions, with either $W$ or $-W$, according to the
coin tossing prescription, with the first laminar region beginning at time $%
t=t_{0}$. Let us imagine, to facilitate the discussion of this section, that
the theoretical sequence is observable, even if in practice it is not. If we
begin the observation process at the same time when the theoretical sequence
is generated, the result of our observation yields the waiting time
distribution of Eq. (\ref{tsallis}). If the observation of the theoretical
sequence begins at a given time $t_{1}>t_{0}$, the distribution of the
waiting times before the first exit from the laminar region, denoted by $%
\psi _{t_{1},t_{0}}(t)$, will not coincide with $\psi (t)$. This is a
consequence of the first laminar region observed having begun at any time
between $t_{1}$ and $t_{0}$. Thus, the resulting waiting time will be, in
general, shorter than the real sojourn time generated by $\psi (\tau )$. In
the Poisson case this shortening of the time does not have any effect on the
shape of $\psi _{t_{1},t_{0}}(t)$, which remains identical to $\psi (\tau )$%
. In the non-Poisson case, on the contrary, delaying the process of
observation does influence the shape of $\psi _{t_{1},t_{0}}(t)$ causing it
to depart from the form of $\psi (\tau )$ \cite{gerardone,barkai}.

Let us now address the problem of building up the aging 
correlation function of $\xi (t)$. We study the correlation between $\xi
(t_{2})$ and $\xi (t_{1})$, with the condition that $t_{2}>t_{1}>t_{0}$; $%
t_{0}$ being the time at which the laminar region begins. We solve this
problem in two steps. In the first step we define the correlation
function $A^{(t_{0})}(t_{2}-t_{1})$, without requiring that the laminar
region begins at $t=t_{0}$, but that it in fact begins at a time intermediate between
$t_1$ and $t_0$. 
This corresponds to stating that $A^{(t_{0})}(t_{2}-t_{1})$ is a
correlation function of undefined age, \emph{younger}, though,
than the $(t_{1}-t_{0})$-old correlation function. In the second step we
set the additional condition that the laminar regions begin at
$t=t_{0}$, and we give the prescription to determine the
correlation function $\Phi^{(t_{0})}_{\xi}$, a notation denoting in
fact the $(t_{1}-t_{0})$-old correlation function.
The latter aging
correlation function fits the earlier definition of $\psi
_{t_{1},t_{0}}(t)$. The corresponding analytical expression will make
it possible to establish the effect of aging on the phenomenon, namely
the effect of moving both $t_{2}$ and $t_{1}$ away from $t_{0}$ as
well as the more traditional effect of increasing the distance between
$t_{2}$ and $t_{1}$.

Note that the first correlation function is given by 
\begin{equation}
A^{(t_{0})}(t_{2}-t_{1})=p(A|\overline{B}).  \label{firstequation}
\end{equation}
This identification of $A^{(t_{0})}(t_{2}-t_{1})$ is consistent because we
define $A$ by the condition that both $t_{1}$ and $t_{2}$ belong to the same
laminar region, while $\overline{B}$ is defined by the condition that $t_{0}$
does not belong to the same laminar region as $t_{1}$. Of course, with this
interpretation $B$ is defined by the condition that $t_{0}$ belongs to the
same laminar region as $t_{1}$.

The conventional correlation function is the probability that $t_{1}$
and $t_{2}$ belong to the same laminar region, and thus is the probabilty
that property $A$ occurs, so we can write the second equality 
\begin{equation}
\Phi _{\xi }(t_{2}-t_{1})=p(A).  \label{secondequation}
\end{equation}
The probability that $A$ and $B$ take place in the same laminar region
allows us to write, in terms of the correlation function, 
\begin{equation}
\Phi _{\xi }(t_{2}-t_{0})=p(A,B).  \label{thirdequation}
\end{equation}
In fact, this is the probability that $t_{2}$ and $t_{0}$ belong to the same
laminar region, and, thanks to the time ordering $t_{2}>t_{1}>t_{0}$, this
is equal to the probability that both $A$ and $B$ occur. Finally, the
probability that $t_{1}$ and $t_{0}$ belong to the same laminar region,
namely, the probability that the property $B$ applies, enables us to write 
\begin{equation}
\Phi _{\xi }(t_{1}-t_{0})=p(B).  \label{fourthequation}
\end{equation}
At this stage, to express $A^{(t_{0})}(t_{2}-t_{1})$ in terms of more
familiar correlation functions we insert Eq. (\ref{noevent}) into Eq.(%
\ref{firstequation}), which yields 
\begin{equation}
A^{(t_{0})}(t_{2}-t_{1})=\frac{\Phi _{\xi }(t_{2}-t_{1})-\Phi _{\xi
}(t_{2}-t_{0})}{1-\Phi _{\xi }(t_{1}-t_{0})}.  \label{second}
\end{equation}
It is easy to show that in the Poisson case Eq.(\ref{second}) reduces to 
\begin{equation}
A^{(t_{0})}(t_{2}-t_{1})=\Phi _{\xi }(t_{2}-t_{1}),  \label{poisson}
\end{equation}
independently of $t_{0}.$

Now let us take the second step, and explicitly evaluate $\Phi _{\xi
}^{(t_{0})}(t_{2}-t_{1})$. This aging correlation function is the sum of
two probabilities. The first contribution is the probabiliy that no event
occurs between $t_{0}$ and $t_{2}$, thereby ensuring that $t_{1}$ and $t_{2}$
belong to the same laminar region. The second contribution is the
probability that an arbitrary number of events occurred between $t_{0}$ and $%
t_{2}$. Note that the laminar region beginning at $t=t_{0}$ implies that at
this time a random event occurs, which is in fact, the beginning of the
laminar region. As stated a number of time earlier, at the beginning of any
laminar region, we toss a coin, to decide the sign of the laminar region.
This is the random event that makes it possible for us to express $\Phi
_{\xi }^{(t_{0})}(t_{2}-t_{1})$ as follows 
\[
\Phi _{\xi }^{(t_{0})}(t_{2}-t_{1})=\Psi (t_{2}-t_{0})
\]
\begin{equation}
+(1-\Psi (t_{1}-t_{0}))%
\frac{\Phi _{\xi }(t_{2}-t_{1})-\Phi _{\xi }(t_{2}-t_{0})}{1-\Phi _{\xi
}(t_{1}-t_{0})}.  \label{paoloaisagenius}
\end{equation}
In Eq. (\ref{paoloaisagenius}) we have used the conventional notation of the
CTRW formalism \cite{montroll},

\begin{equation}
\Psi (t)\equiv \int_{t}^{\infty }dt^{\prime }\psi (t^{\prime }),
\label{wait}
\end{equation}
where $\psi (t)$ is the waiting time distribution of Eq.(\ref{tsallis}).
Montroll and Weiss \cite{montroll} make the implicit assumption that the
laminar region begins at $t=0$. Thus, $\Psi (t)$ is the probability that no
event occurs up to time $t,$ after the random event occurs at time $t=0$.
Here we replace the initiation time $t=0$ with $t=t_{0}$. Thus, $\Psi
(t_{2}-t_{0})$ is the probability that no random event occurs between $t_{0}$
and $t_{2}$, as required. The second term in Eq.(\ref{paoloaisagenius}) is
the product of the probability that one or more events occurred between $%
t_{1}$ and $t_{0}$, given the fact that $t_{2}$ and $t_{1}$ are in the same
laminar region and $t_{0}$ is not.

We note that Eq. (\ref{paoloaisagenius}) interrelates factorability and
aging and consequently is the most relevant expression for our discussion.
The importance of this result can be made transparent by going back to the
discussion in Section 2. Eq. (\ref{centralresult}), the expression for the
fourth-order correlation function, can be reexpressed as, using Eq.(\ref
{second}), 
\begin{eqnarray}
\frac{\langle \xi (t_{1})\xi (t_{2})\xi (t_{3})\xi (t_{4})\rangle }{\langle
\xi ^{2}\rangle ^{2}} =\Phi _{\xi }(t_{4}-t_{1})  \nonumber \\
+(\Phi _{\xi }(t_{2}-t_{1})-\Phi _{\xi
}(t_{3}-t_{1}))A^{(t_{2})}(t_{4}-t_{3}). &&  \label{newform}
\end{eqnarray}
As pointed earlier, in the Poisson case, see Eq. (\ref{poisson}), 
\begin{equation}
A^{(t_{2})}(t_{4}-t_{3})=\Phi _{\xi }(t_{4}-t_{3}),  \label{newcase}
\end{equation}
independently of $t_{2}$. By inserting Eq.(\ref{newcase}) into Eq. (%
\ref{newform}), and noting that $\Phi _{\xi }(t_{4}-t_{1})=$ $\Phi _{\xi
}(t_{4}-t_{3})\Phi _{\xi }(t_{3}-t_{1}),$ we see immediately that the DF
condition is recovered:

\begin{equation}
\frac{\langle \xi (t_{1})\xi (t_{2})\xi (t_{3})\xi (t_{4})\rangle }{\langle
\xi ^{2}\rangle ^{2}}=\Phi _{\xi }(t_{2}-t_{1})\Phi _{\xi }(t_{4}-t_{3}).
\label{final}
\end{equation}
Thus, we have established that the breakdown of the DF condition and aging
are interrelated. In fact, annihilating the aging property has the effect of
reestablishing the DF property.

\section{Concluding remark}

The equivalence between the trajectory and density pictures of physical
phenomena is one of the major tenants of modern physics. It therefore came
as quite a suprise when Bologna \textit{et al. }\cite{bolognone} discovered
an inconsistency between these two pictures in the case of non-ordinary
statistical mechanics. The form of the inconsistency had to do with the
derivation of anomalous diffusion of the L\'{e}vy kind, using dichotomous
noise and either CTRW or the generalized central limit theorem. Both of
these approaches use trajectories and not the Liouville-like approach for
densities, such as does GDE. It is a simple matter, using Eq.(\ref{GDE}), to
show that GDE yields a hierarchy of moments $\left\langle x^{2n}\left(
t\right) \right\rangle $ with $n=1,2,...,$ which coincides exactly with the
hierarchy generated by fluctuations $\xi \left( t\right) $ satisfying DF.
This factorization, obtained using the density, contradicts the hierarchy
generated using the trajectories in Section 3. We have limited the analysis
to the fourth-order correlation functions, however, this order is sufficient
to identify the source of the inconsistency between the trajectory and
density pictures as being due to the non-Poisson character of the statistics.

We have also shown that a departure from Poisson statistics has the
effect of introducing a memory into the correlation functions that can
last for an infinitely long time. For dichotomous noise the two-time
correlation function, using either trajectories or densities is
the same, however, higher-order correlations are not the same for
non-Poisson statistics. The deviation from Poisson statistics is
manifest in a dependence of correlations on the difference between the
initiation time and the observation time, that is, on the age of the
system. Age destroys the DF property and may represent a state of
matter intermediate between the dynamic and thermodynamic condition,
mentioned earlier, the Living State of Matter. This eternal state of
nonequilibrium, in which a perturbed phenomena relaxes to, but never
attains, equilbrium, should be contrasted with the Onsager Principle
in which physical systems are assumed to be aged. An aged physical
system is one that has reached equilibrium with a heat bath long
before measurements are taken.

It is evident that to establish a density picture equivalent to the
trajectory picture, in which the time averages and ensemble averages are the
same, in the non-Poisson as well as in the Poisson case, we have to overcome
the limitations of the Liouville-like approaches of Refs. \cite
{lemma,bourret,zwanzig}. This difficult issue calls for further research.
Nevertheless, the merit of the present paper lies in the fact that it has
revealed the violation of the DF property when the statistics of the
underlying process are non-Poisson. DF is a factorization property assumed
for dichotomous noise by researchers in multiple fields, often without the
realization that such factorization is tied to the statistics of the
process. 

It is worth remarking that Eq. (\ref{GDE}), the general diffusion equation,
can also be derived by assuming that the random sequence $\xi \left(
t\right) $ is built up by time-modulating a generating Poisson distribution $%
\psi (\lambda (t),t)=exp(-\lambda (t)t)$ as shown in detail by Bologna
\emph{et al. }\cite{pala}. 
The resulting sequence, however, is not a renewal sequence,
such as found in CTRW. We need to understand why abandoning the Poisson
assumption and adopting a Liouville-like approach leads to physical effects
that are inconsistent with renewal processes. This is a difficult problem
whose solution also requires additional research. It is important to point
out, to avoid any possible confusion, that the GDE is widely used to
describe transport processes (see Refs. \cite{bisquert,katja}, for some
recent papers). However, these papers refer to subdiffusion, a physical
condition where the correlation function of the fluctuation $\xi $ cannot be
defined; not even in the non-stationary sense of Section 3. The discussion
herein focuses on superdiffision and addresses the problem of computing
high-order correlations for renewal process with non-exponential waiting
time distributions. The solution to this problem is given by Eq. (\ref
{centralresult}), however this crucial property has not yet been obtained
using Liouville-like methods \cite{lemma,bourret,zwanzig}.


In conclusion, by means of the conditional probability formalism, we
have found the exact expression for the fourth-order correlation
function, and we have shown that in the non-Poisson case, this
expression violates the DF condition. We have also established a close
connection between the DF breakdown and aging. In the case where $\mu>
2$ the aged condition is possible. However, if $\mu < 3$, the aging
condition lasts forever. We see, in fact, from Eq. (\ref{geisel}) that in this
case the correlation function $\Phi_{\xi}(t)$ is an inverse power
law with index $\mu-2$. Thus, it takes an infinitely long time
for the age-dependent correlation function
Eq. (\ref{paoloaisagenius}) 
to become
stationary. This is a remarkable result, which challenges the
traditional treatments of such stochastic dynamical processes based on
the generalized master equation (GME). The analysis of the GME based
on these results will be taken up elsewhere.

\emph{Acknowledgements}. PG thankfully aknowledges the Army Research Office
for financial support through Grant DAAD19-02-1-0037

\end{document}